# Integrated Nested Laplace Approximation for Bayesian Nonparametric Phylodynamics

**Julia A. Palacios**
Department of Statistics
University of Washington
jpalacio@uw.edu

**Vladimir N. Minin**
Department of Statistics
University of Washington
vminin@uw.edu

## Abstract

The goal of phylodynamics, an area on the intersection of phylogenetics and population genetics, is to reconstruct population size dynamics from genetic data. Recently, a series of nonparametric Bayesian methods have been proposed for such demographic reconstructions. These methods rely on prior specifications based on Gaussian processes and proceed by approximating the posterior distribution of population size trajectories via Markov chain Monte Carlo (MCMC) methods. In this paper, we adapt an integrated nested Laplace approximation (INLA), a recently proposed approximate Bayesian inference for latent Gaussian models, to the estimation of population size trajectories. We show that when a genealogy of sampled individuals can be reliably estimated from genetic data, INLA enjoys high accuracy and can replace MCMC entirely. We demonstrate significant computational efficiency over the state-of-the-art MCMC methods. We illustrate INLA-based population size inference using simulations and genealogies of hepatitis C and human influenza viruses.

## 1 INTRODUCTION

Estimation of population size dynamics from molecular data is a fundamental task in ecology and public health. Since population size fluctuations affect the variability of population gene frequencies, current molecular sequence data provide information about the past population size trajectory. Such indirect inference is particularly useful in retrospective studies, where assessing past population sizes via sampling or fossil records is impossible. For example, RNA samples of hepatitis C virus (HCV) obtained in 1993 were sufficient to estimate the dynamics of HCV infections in Egypt from 1895 to 1993 (Pybus et al., 2003); and ancient and modern musk ox mitochondrial DNA samples, dated from 56,900 radiocarbon years old to contemporaneous, allowed for estimation of musk ox population dynamics throughout the late Pleistocene to the present (Campos et al., 2010).

Molecular sequence data of individuals sampled at a single time point (*isochronous* sampling) or at different points in time (*heterochronous* sampling) are related to each other via, a usually unknown, genealogical relationship. A genealogy is a rooted bifurcating tree that describes the ancestral relationships of the sampled individuals (left upper box in Figure 1). In the genealogy, each internal node indicates that the two lineages met a common ancestor. Such events are called *coalescent events*, and these events occur at *coalescent times*.

Kingman's coalescent (Kingman, 1982) is a probability model that describes a stochastic process of generating a genealogy of a random sample of molecular sequences given the effective population size (Nordborg, 2001; Hein et al., 2005). The original formulation, that considered only a constant population size, was later generalized to a variable population size (Slatkin and Hudson, 1991; Griffiths and Tavaré, 1994). Statistically, the coalescent model was an important advance, because it allowed for likelihood-based inference of population dynamics.

Many coalescent-based methods for estimation of effective population size trajectories have been developed over the last 10 years. For a recent review see (Ho and Shapiro, 2011). Some methods assume that a fixed genealogy is available (Fu, 1994; Pybus et al., 2000) and others may or may not consider the genealogical uncertainty and can produce estimates of population size trajectories from a fixed genealogy or directly from molecular data (Kuhner et al., 1995; Drummond et al., 2002, 2005; Minin et al., 2008). Felsenstein (1992) showed that likelihood-based methods that ac-count for genealogical uncertainty are statistically the most efficient. However, all methods that incorporate genealogical uncertainty in population size dynamics reconstruction integrate over the space of genealogies using Markov chain Monte Carlo (MCMC). Such MCMC sampling of genealogies is computationally expensive. Sometimes, a single genealogy estimated from sequences that contain sufficient phylogenetic information is enough to estimate population trajectories accurately (Pybus et al., 2000; Minin et al., 2008). In this paper, we are interested in providing a fast estimation of population size trajectories from a fixed genealogy.

Some coalescent-based methods assume a simple parametric form of the population size trajectory (e.g., exponential or logistic growth), allowing the model parameters to be estimated by maximum likelihood or Bayesian methods. However, more flexible nonparametric methods are preferable for populations with poorly understood population dynamics, where it may be difficult to justify a simple parametric form of the population size trajectory. In fact, all recently developed methods rely on Bayesian nonparametric techniques to perform inference (Opgen-Rhein et al., 2005; Drummond et al., 2005; Heled and Drummond, 2008; Minin et al., 2008; Palacios and Minin, 2011). A common characteristic of most of these methods is the assumption of a piece-wise linear trajectory of effective population sizes and the possibility of the number of parameters growing with the number of samples. Bayesian skyline methods (Drummond et al., 2005; Heled and Drummond, 2008) and Opgen-Rhein et al. (2005) use multiple change point models to estimate population trajectories in a Bayesian framework. The method of Opgen-Rhein et al. (2005) is implemented only for a fixed genealogy. Recently, Bayesian nonparametric approaches that rely on Gaussian processes have been successfully implemented (Minin et al., 2008; Palacios and Minin, 2011). These methods model the effective population size as a function of a Gaussian process (GP) *a priori*, providing more flexible priors than previous Bayesian nonparametric methods.

GP-based models use MCMC methods to perform Bayesian inference. We show that when the genealogy remains fixed, these models fall into a general class of latent Gaussian models, for which integrated nested Laplace approximation (INLA) can be used to perform computationally efficient approximate Bayesian inference (Rue et al., 2009; Illian et al., 2012). Here, we adapt the INLA methodology to the estimation of population size trajectories and replace MCMC entirely. Our approximation is accurate and much faster than MCMC, while still providing the benefits of the Gaus-sian process-based Bayesian nonparametric approach. We illustrate the performance of our method with simulated and two real data sets.

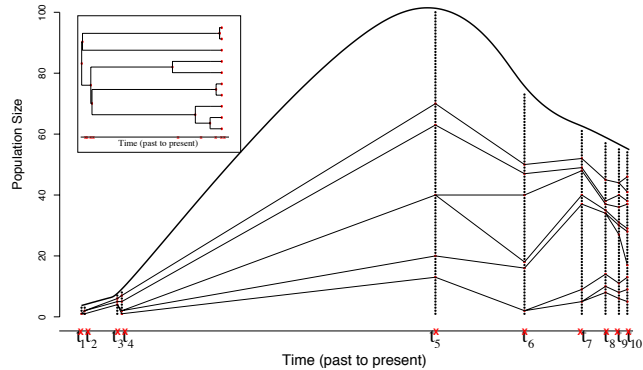

Figure 1: Example of a genealogy of 10 individuals randomly sampled at time $t_{10}$ (red circles) from the population depicted as black circles. When we follow the ancestry of the samples back in time, two of those lineages coalesce at time $t_9$, the rest of the lineages continue to coalesce until the time to the most recent common ancestor of the sample at time $t_1$. The population size trajectory is shown as the solid black curve. When the population size is large (around $t_5$), for any pair of lineages that exist at time $t_5$ (red circles at $t_5$), it takes longer to meet their most recent common ancestor ($t_4$). The upper left box shows the genealogy reconstructed by following the ancestry of the 10 sampled individuals. The genealogy in the upper left corner is the aligned representation of the genealogy depicted in the main plot.

## 2 COALESCENT BACKGROUND

We assume that a genealogy with time measured in units of generations is available. Let $t_n = 0$ denote the present time when all $n$ available sequences are sampled (*isochronous*) and let $t_n = 0 < t_{n-1} < ... < t_1$ denote the coalescent times of lineages in the genealogy. Figure 1 depicts an example of such a genealogy with time going backwards, that is, the first coalescent time occurred $t_{n-1}$ generations ago and all the samples meet the common ancestor $t_1$ generations ago. Let $N_e(t)$ denote the time evolution of the effective population size as we move into the past. Then, the conditional density of the coalescent time $t_{k-1}$, given the previous coalescent time $t_k$, takes the following form:

$$P[t_{k-1}|t_k, N_e(t)] = \frac{C_k \exp\left[-\int_{t_k}^{t_{k-1}} \frac{C_k}{N_e(t)} dt\right]}{N_e(t_{k-1})}, \quad (1)$$

where $C_k = \binom{k}{2}$ is the coalescent factor that depends on the number of lineages $k = 2, ..., n$, meaning that the density for the next coalescent time is quadratic in the number of lineages and inversely proportional to the effective population size. The larger the popu-lation size, the more genetic variability is in the population and hence, the longer it takes for two lineages to coalesce. The larger the number of lineages, the faster two of them meet their common ancestor. Figure 1 shows an example of a population that experiences growth and then a decay in population size. In this case, no pair of lineages coalesces between times $t_4$ and $t_5$, because the population is large during this time interval, while it takes little time for a pair of lineages to find their common ancestor after time $t_4$, when the population size becomes very small.

The *heterochronous* coalescent arises when samples of sequences are collected at different times. The conditional density of a coalescent time $t_{k-1}$ is slightly different than Eq. 1 since it takes into account the fact that the number of lineages at each time point depends not only on the number of coalescent events (in which case, the number of lineages decreases by one each time), but also on the new samples incorporated into the analysis at any time after the last coalescent time $t_k$. The details of the heterochronous coalescent are omitted for brevity, however, all methods described here have been implemented to incorporate heterochronous sampling. See (Felsenstein and Rodrigo, 1999) and (Drummond et al., 2002) for a more detailed account of heterochronous sampling.

Under this coalescent-based framework, we ignore the effects of population structure, recombination and selection (Nordborg, 2001). The parameter of interest, the effective population size, can be used to approximate census population size by knowing the generation time in calendar units and the population variability in the number of offspring. The latter quantity might be difficult to know *a priori*, however, sometimes it suffices to analyze an arbitrarily rescaled population size trajectory, assuming the variability in the number of offspring remains constant.

## 2.1 ESTIMATION OF $N_e(t)$ USING A DISCRETE-TIME GMRF

There are two approaches to estimation of effective population size trajectories that use Gaussian processes. The first approach, developed by Minin et al. (2008), assumes *a priori* that given a genealogy, the effective population size trajectory is a piecewise constant trajectory with change points (knots) placed at coalescent times. That is,

$$N_e(t) = \sum_{k=2}^{n} \exp(\gamma_k) 1_{(t_k, t_{k-1}]}(t), \quad (2)$$

where

$$\boldsymbol{\gamma} = (\gamma_2, ..., \gamma_k) \sim MVN\left(\mathbf{0}, (\tau \mathbf{Q})^{-1}\right) \text{ and}$$

$$1_{(t_k, t_{k-1}]}(t) = \begin{cases} 1 & \text{if } t \in (t_k, t_{k-1}], \\ 0 & \text{otherwise}. \end{cases}$$

More specifically, *a priori* $\boldsymbol{\gamma}$ is assumed to be an intrinsic Gaussian Markov random field (GMRF) on a chain graph connecting nodes 2 through $n$. Minin et al. (2008) used a random walk of the first order (rw1) on an irregular grid of mid-points of inter-coalescent time intervals. For this reason, we refer to this method here as the coalescent grid Gaussian process (CGGP). The random walk construction implies that matrix $\mathbf{Q}$ is tridiagonal and positive semidefinite (hence the intrinsic GMRF). See (Rue and Held, 2005) for background on GMRFs. The precision parameter $\tau$ has a Gamma prior distribution with $\alpha = \beta = 0.001$. The authors estimate $\boldsymbol{\gamma}$ and $\tau$ by MCMC sampling from the posterior distribution of these parameters. The estimated trajectory and the corresponding uncertainty are reported in the form of pointwise posterior medians and 95% Bayesian credible intervals (BCIs) obtained from the MCMC samples.

## 2.2 ESTIMATION OF $N_e(t)$ USING A CONTINUOUS-TIME GP

Instead of modelling $N_e(t)$ as a piecewise continuous function *a priori*, Palacios and Minin (2011) propose a more flexible prior specification and place a transformed Gaussian process prior on $N_e(t)$. The transformation is a sigmoidal function with a lower bound. This particular transformation is required by the authors in order to perform exact posterior inference via a data augmentation scheme, which is similar to the work of Adams et al. (2009). However, a log-Gaussian transformation using a finely discretized Gaussian process, in principle, would produce similar results (Møller et al., 1998; Adams et al., 2009).

### 2.2.1 EXACT POSTERIOR INFERENCE WITH GP

Palacios and Minin (2011) place the following prior on $N_e(t)$:

$$N_e(t) = \left(\frac{\lambda}{1 + \exp[-\gamma(t)]}\right)^{-1}, \quad (3)$$

where

$$\gamma(t) \sim \mathcal{GP}(0, C) \quad (4)$$

and $\mathcal{GP}(0, C)$ denotes a Gaussian process with mean function 0 and covariance function $C$. A Gaussian process restricted to finite data is a multivariate Gaussian distribution. That is, $\gamma(t_1), ..., \gamma(t_B) \sim MVN(\mathbf{0}, \boldsymbol{\Sigma})$. *A priori*, $1/N_e(t)$ is a sigmoidal Gaussian process, a scaled logistic function of a Gaussian process which range is restricted to lie in $[0, \lambda]$; $\lambda$ is a positive constant hyperparameter, inverse of which serves as a lower bound of $N_e(t)$ (Adams et al., 2009). The likelihood function is the product of the conditional densities in Eq. 1 and involves integration of $N_e(t)$, that under the $\mathcal{GP}$ assumption, is intractable. The authors, following earlier work by Adams et al. (2009) on Poisson processes, do inference assuming an augmented data likelihood which allows to bypass intractability in the likelihood. The authors implement their method for the Brownian motion $\mathcal{GP}$ with a precision parameter $\tau$. They place a Gamma prior distribution on the precision hyperparameter $\tau$ with $\alpha = \beta = 0.001$ and a mixture of uniform and exponential distributions on an upper bound of $1/N_e(t)$ (or equivalently, a lower bound on $N_e(t)$) as follows:

$$P(\lambda) = \epsilon \frac{1}{\hat{\lambda}} I_{\{\lambda < \hat{\lambda}\}} + (1-\epsilon) \frac{1}{\hat{\lambda}} e^{-\frac{1}{\hat{\lambda}}(\lambda - \hat{\lambda})} I_{\{\lambda \geq \hat{\lambda}\}}, \quad (5)$$

where $\epsilon > 0$ is a mixing proportion and $\hat{\lambda}$ is our best guess of the upper bound, possibly obtained from previous studies. The authors estimate $\tau$ and $N_e(t)$, or equivalently, $\tau$, $\gamma(t)$ and $\lambda$ by MCMC sampling from the posterior distribution of these parameters. The estimated trajectory and the corresponding uncertainty are reported in the form of the pointwise posterior medians and 95% BCIs evaluated at a grid of points $\{s_1, ..., s_B\}$ obtained from the MCMC samples. This grid can be made as fine as necessary after the MCMC is finished. The values of $\{\gamma(s_1), \gamma(s_2), ..., \gamma(s_B)\}$ are obtained via the $\mathcal{GP}$ predictive distribution conditioning on the values of each iteration. This method will be referred to as exact Gaussian process (EGP).

#### 2.2.2 DISCRETIZED CONTINUOUS-TIME GP

The continuous-time version of the prior specified in Eq. 2, is

$$N_e(t) = \exp[\gamma(t)], \quad (6)$$

where $\gamma(t)$ is the Gaussian process described in Eq. 4. However, for the same reason described in section 2.2.1, the likelihood function becomes intractable. Palacios and Minin (2011) showed that estimation of the effective population size is analogous to the estimation of an inhomogeneous intensity of a point process. In this context, and under the prior described in Eq. 6, estimation of $N_e(t)$ is computationally equivalent to the estimation of the intensity function of a Log-Gaussian Cox process (Møller et al., 1998). In a Log-Gaussian Cox process, the likelihood is commonly approximated by discretization. The approximation method proceeds by constructing a fine regular grid $\{s_1, ..., s_B\}$ over the observation window and approximate

$$\int \frac{dt}{N_e(t)} = \int \exp[-\gamma(t)] dt, \quad (7)$$

by

$$\sum_{j=2}^{B} \exp\left(-\gamma_j^*\right) \Delta, \quad (8)$$

where $\Delta$ is the distance between grid points, and $\gamma_j^*$ is a representative value of $\gamma(t)$ in the interval $(s_{j-1}, s_j)$, usually $\gamma((s_j - s_{j-1})/2)$. Note that if the Gaussian process is a Brownian motion process, this approximation is similar to the CGGP method described in section 2.1. The difference is in the construction of the grid. In the CGGP method, the grid is irregular and determined by the coalescent times. For this reason, we call approximation (8) a regular grid Gaussian process (RGGP).

## 3 INTEGRATED NESTED LAPLACE APPROXIMATION

INLA provides fast and accurate Bayesian approximation to posterior marginals in *latent Gaussian models* (Rue et al., 2009). Latent Gaussian models are a wide class of hierarchical models in which the response variables $\mathbf{y} = (y_1, \ldots, y_n)$ are assumed to be conditionally independent given some latent parameters $\boldsymbol{\eta} = (\eta_1, \ldots, \eta_n)$ and other parameters $\boldsymbol{\theta}_1$. The second hierarchical level corresponds to specifying $\boldsymbol{\eta}$ as a function of a GMRF $\mathbf{x} = (x_1, \ldots, x_n)$ with a precision matrix $\mathbf{Q}$ and hyperparameters $\boldsymbol{\theta}_2$, and the third and last hierarchical stage corresponds to prior specifications for the hyperparameters $\boldsymbol{\theta} = (\boldsymbol{\theta}_1, \boldsymbol{\theta}_2)$ Formally,

$$\pi(\mathbf{y}|\boldsymbol{\eta}, \boldsymbol{\theta}_1) = \prod_j \pi(y_j|\eta_j(x_j), \boldsymbol{\theta}_1), \quad (9)$$

$$\mathbf{x} \sim MVN(\mathbf{0}, \mathbf{Q}^{-1}(\boldsymbol{\theta}_2)), \quad (10)$$

and

$$\boldsymbol{\theta} \sim P(\boldsymbol{\theta}). \quad (11)$$

An interface in R, called INLA, implements a wide variety of likelihoods (Eq. 9), link functions ($\boldsymbol{\eta}$) and GMRFs (Eq. 10), including the Poisson likelihood model for each observed value of $y_j$ (not necessarily the same for every $y_j$) with a logarithmic additive link function and random walk of first order as a GMRF. See www.r-inla.org for documentation.

The coalescent with variable population size (Eq. 1), together with the GMRF prior specification (Eq. 2) falls into the latent Gaussian model class, so INLA can be implemented for these coalescent models. In the case of the continuously specified $\mathcal{GP}$ (section 2.2), the approximate posterior method described in Section 2.2.2 (RGGP) also falls into the latent Gaussian model class.

## 3.1 INLA FOR PHYLODYNAMICS

Although INLA is implemented for a wide variety of latent Gaussian models, we will only describe the main steps of the approximation for posterior inference of effective population size trajectories. A typical summary of the posterior distribution of the effective population size trajectory, $N_e(t)$, is described by posterior medians and 95% BCIs evaluated pointwise on a grid of time points. These values can be obtained from the posterior marginals of the population trajectory on the grid. For the CGGP model described in section 2.1, we then wish to obtain the posterior marginals

$$\Pr(\gamma_i|\mathbf{t}) = \int_0^\infty \Pr(\gamma_i|\tau,\mathbf{t})\Pr(\tau|\mathbf{t})d\tau, i = 2,..,n \quad (12)$$

and

$$\Pr(\tau|\mathbf{t}), \quad (13)$$

where $\mathbf{t}$ denotes the vector of coalescent times. A nested procedure is used to construct approximations of $\Pr(\gamma_i|\tau,\mathbf{t})$ and $\Pr(\tau|\mathbf{t})$ first and then numerically integrate out $\tau$ to arrive at $\Pr(\gamma_i|\mathbf{t})$. The approximation of the marginal of $\tau$ is

$$\tilde{\Pr}(\tau|\mathbf{t}) \propto \left.\frac{\Pr(\boldsymbol{\gamma},\tau,\mathbf{t})}{\tilde{\Pr}_\mathbf{G}(\boldsymbol{\gamma}|\tau,\mathbf{t})}\right|_{\boldsymbol{\gamma}^*(\tau)}, \quad (14)$$

where $\boldsymbol{\gamma}^*(\tau)$ is the mode of the full conditional $\Pr(\boldsymbol{\gamma}|\tau,\mathbf{t})$, obtained using the Newton-Raphson algorithm, and $\tilde{\Pr}_G(\boldsymbol{\gamma}|\tau,\mathbf{t})$ is the Gaussian approximation of this full conditional constructed via a Taylor expansion around $\boldsymbol{\gamma}^*(\tau)$. The resulting $\tilde{\Pr}_G(\boldsymbol{\gamma}|\tau,\mathbf{t})$ is a Gaussian distribution with mean $\boldsymbol{\gamma}^*$ and precision matrix $\mathbf{Q}+\mathrm{diag}(\mathbf{c})$, where $\mathbf{Q}$ is the prior precision matrix of the GMRF $\boldsymbol{\gamma}$ and a vector $\mathbf{c}$ consists of the second order Taylor series coefficients.

The approximation to the full conditional $\Pr(\gamma_i|\tau,\mathbf{t})$ is the following:

$$\tilde{\Pr}(\gamma_i|\tau,\mathbf{t}) \propto \left.\frac{\Pr(\boldsymbol{\gamma},\tau,\mathbf{t})}{\tilde{\Pr}_\mathbf{G}(\boldsymbol{\gamma}_{-i}|\tau,\mathbf{t})}\right|_{\boldsymbol{\gamma}^*_{-i}}, \quad (15)$$

where $\boldsymbol{\gamma}^*_{-i} = \mathrm{E}_G(\boldsymbol{\gamma}_{-i}|\gamma_i,\tau,\mathbf{t})$ and $\tilde{\Pr}_G(\boldsymbol{\gamma}_{-i}|\tau,\mathbf{t})$ are derived from $\tilde{\Pr}_G(\boldsymbol{\gamma}|\tau,\mathbf{t})$.

For the continuously specified GP approximation described in section 2.2.2, the INLA approximation is, in essence the same, but the GMRF is placed at the mid-points of a finer and regular grid. In this case, there are two levels of approximation, one level corresponding to the likelihood discretization and another level corresponding to the approximation of marginal posterior distributions of model parameters.

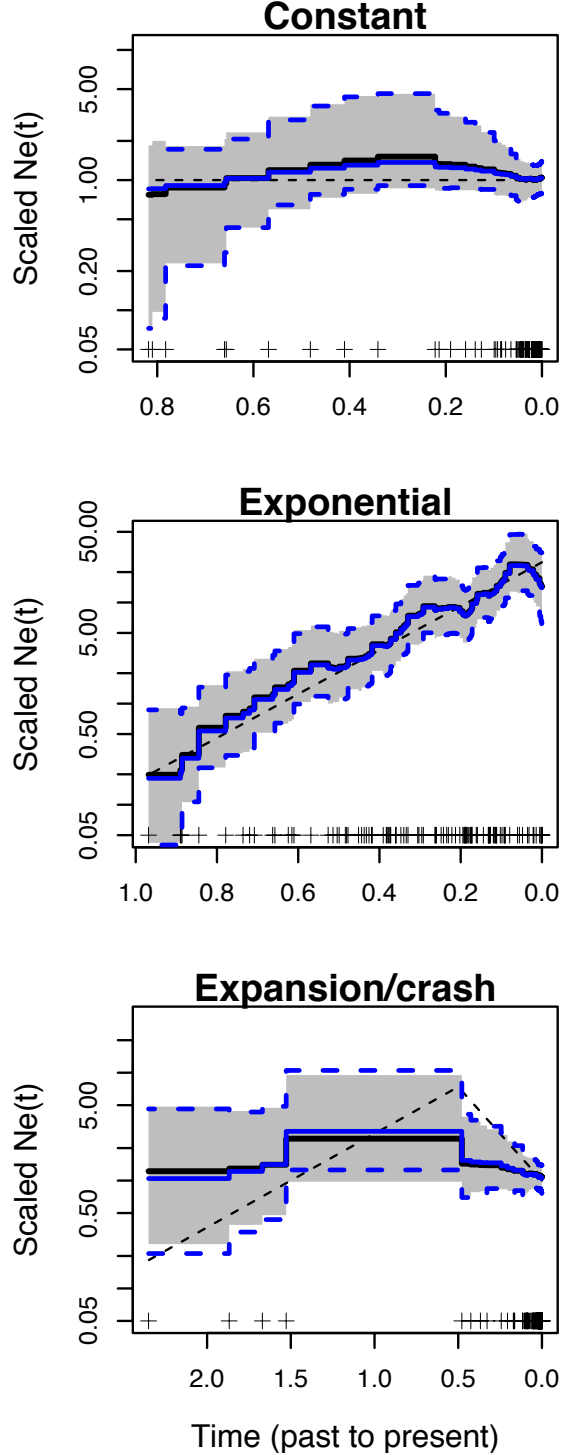

Figure 2: INLA vs MCMC for CGGP: Simulated data under the constant population size (first row), exponential growth (second row) and expansion followed by a crash (third row). The true trajectories are represented by black dashed lines. We show posterior medians estimated with MCMC sampling (solid black lines) and 95% BCIs estimated with MCMC (gray shaded areas). Posterior medians obtained using INLA are denoted by solid blue lines and INLA 95% BCIs are shown as dashed blue lines.

# 4 RESULTS

## 4.1 SIMULATED DATA

We compare INLA and MCMC approaches for the models described in sections 2.1 and 2.2. We simulate three genealogies relating $n = 100$ individuals under the following demographic scenarios:

1. Constant population size trajectory: $N_e(t) = 1$.

2. Exponential growth: $N_e(t) = 25e^{-5t}$.

3. Population expansion followed by a crash:

$$N_e(t) = \begin{cases} e^{4t} & t \in [0, 0.5], \\ e^{-2t+3} & t \in (0.5, \infty). \end{cases} \quad (16)$$

Figure 2 shows the log effective population size trajectories recovered for the three scenarios under the CGGP model using the MCMC approach (black lines and gray shaded areas) and the INLA approach (blue dark lines and blue dashed lines). In all the cases, the INLA approximation is very close to the results obtained using MCMC.

Figure 3 shows the log effective population size trajectories recovered for the same three scenarios for the continuously specified GP. In this case, the comparison is not entirely fair because we are comparing the exact MCMC method (EGP) with the doubly approximated INLA on the RGGP model. Nevertheless, both estimations look very similar for the last two cases (exponential growth and expansion followed by crash). In all cases, INLA results are very similar to the results for the CGGP model and the difference between the MCMC method and INLA methods in the constant trajectory example could be an artifact of the likelihood approximation and the convergence of the MCMC method. However, a more likely explanation is poor approximation of the marginal posterior of the Brownian motion precision, $\tau$, by INLA. Indeed, when we examined MCMC-based and INLA-based marginal posteriors of $\tau$, we found that the two marginals did not agree at all.

## 4.2 HEPATITIS C VIRUS IN EGYPT

We analyze a genealogy estimated from 63 HCV E1 sequences sampled in 1993 in Egypt. This is perhaps the most commonly used dataset for evaluating different methodologies for estimation of population size trajectories. Minin et al. (2008) compared population size trajectories recovered from a single fixed genealogy and from the sequence data directly. The authors show that there is little difference between these two estimation protocols. They argue that in this case genealogical uncertainty does not play a significant role in the estimation of the Egyptian HCV population dynamics.

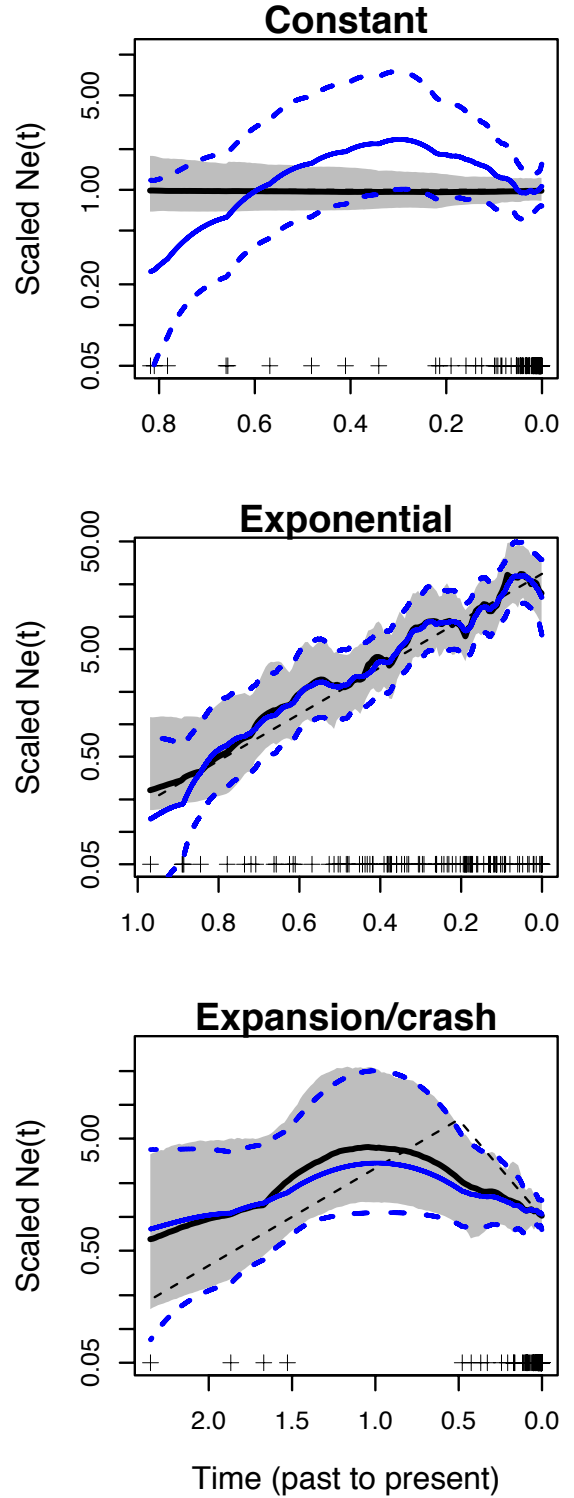

Figure 3: INLA vs MCMC for RGGP and EGP respectively: see Figure 2 for the legend.

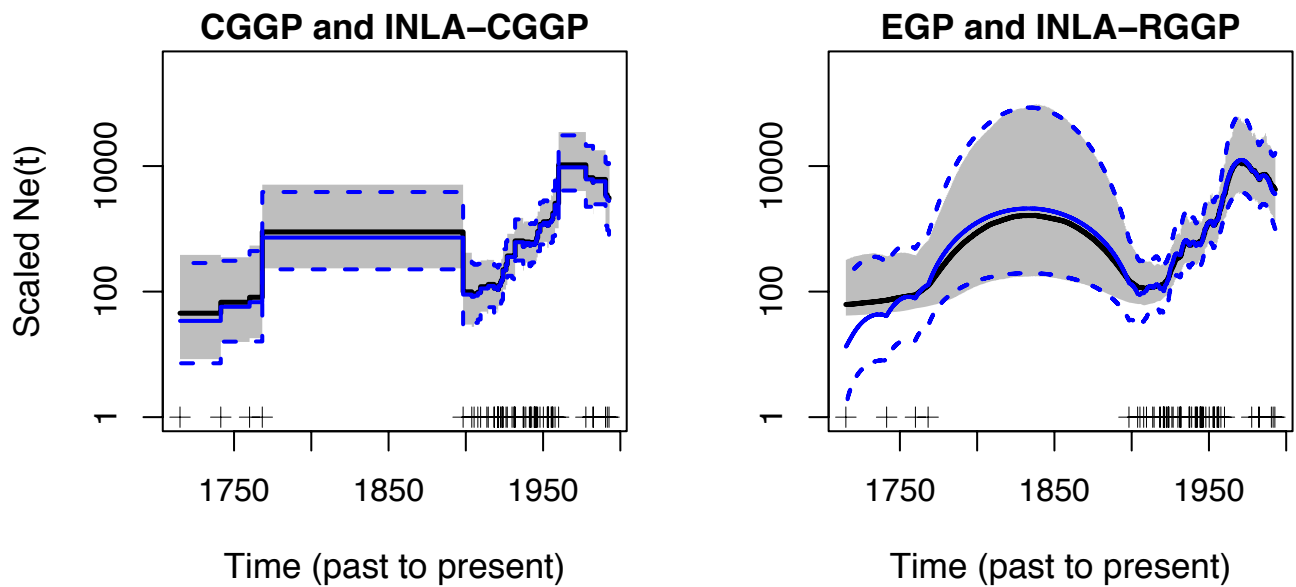

Figure 4: HCV in Egypt. Estimation of the log effective population size trajectories. In both plots, INLA approximations to posterior medians and 95% BCIs are represented by blue solid lines and blue dashed lines respectively. Approximations using MCMC sampling are represented by black solid lines and shaded areas. The left plot shows the results assuming the CGGP model and the right plot shows the result assuming the EGP for the MCMC sampling results and the RGGP model for the INLA approximation.

Figure 4 shows the recovered effective population sizes as black lines and uncertainty as gray shaded areas for the CGGP (left plot) and the EGP (right plot) using MCMC and as blue solid lines and blue dashed lines for the INLA approximation for CGGP (left plot) and RGGP (right plot). In this case, it is remarkable how similar the INLA approximations are to the MCMC results, even for the continuously specified model with the double approximation (INLA-RGGP). In all cases, the known aspects of the HCV epidemic in Egypt are recovered: an exponential growth starting around 1920s and a decline in population size after 1970s (Pybus et al., 2003).

### 4.3 INFLUENZA A VIRUS IN NEW YORK

We analyze a genealogy estimated from 288 H3N2 sequences sampled in New York state from January, 2001 to March, 2005 to estimate population size dynamics of human influenza A in New York. This genealogy has also been analyzed before (Palacios and Minin, 2011) and can be obtained from the authors. The key aspects of the influenza A virus epidemic in temperate regions like New York are the epidemic peaks during winters followed by strong bottlenecks at the end of the winter season. The first plot in Figure 5 shows the recovered population size trajectories assuming the CGGP model. In this case, the MCMC and the INLA approximation deviate from each other substantially, however, the expected peaks during the winter seasons in 2002, 2004 and 2005 are recovered by both methods. The MCMC approach does not recover a peak in the 2003 season, while the INLA approximation resemble more the results from the continuously specified model. INLA and MCMC results are very similar for the continuously specified model (right plot of Figure 5) with the notable differences in 95% BCIs near the time to the most recent common ancestor. This difference again may be an artifact of the double approximation involved.

### 4.4 RUNNING TIMES

The MCMC chains used for the CGGP model have length 1,000,000 with 100,000 of burn-in and generated using the BEAST software (Drummond and Rambaut, 2007; Minin et al., 2008) on a desktop PC. The running times range from 20 minutes to a couple of hours depending on the data. For the INLA approach, results were generated using the R interface INLA on the same computer in less than 2 seconds for all scenarios.

For the continuously specified GP model described in section 2.2, MCMC times are at best as fast as MCMC for the CGGP approach, while the results obtained

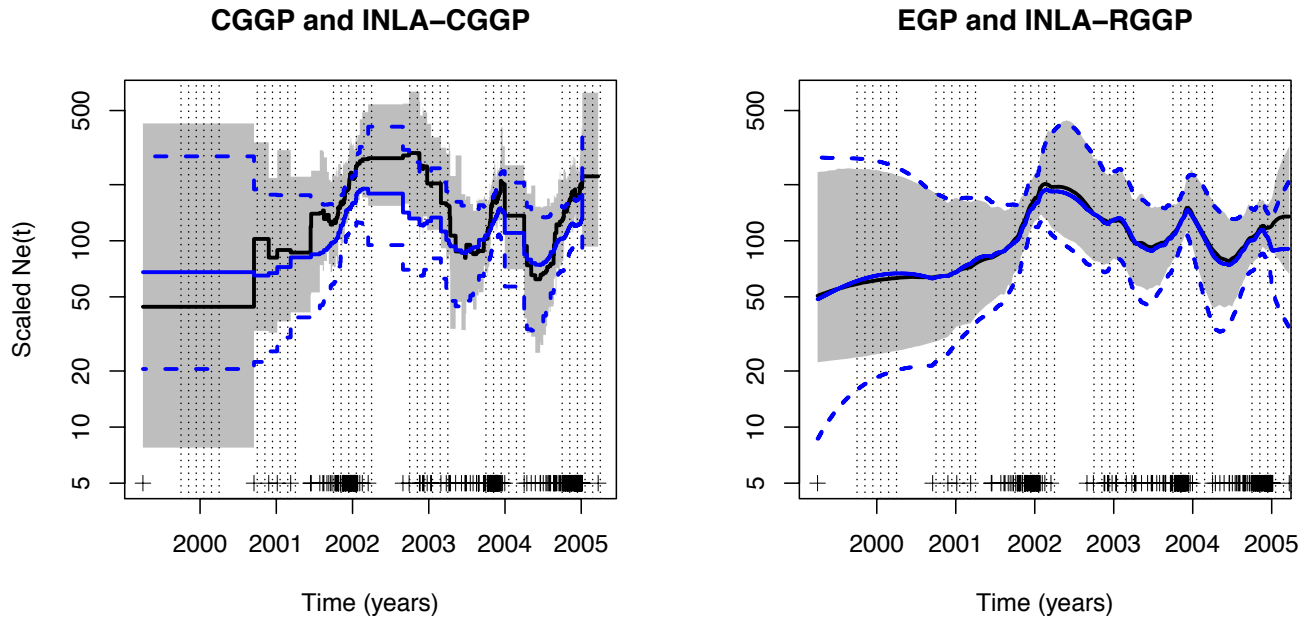

Figure 5: Influenza A in New York. Estimation of the log effective population size trajectories. In both plots, INLA approximations to posterior medians and 95% BCIs are represented by blue solid lines and blue dashed lines respectively. Approximations using MCMC sampling are represented by black solid lines and shaded areas. The left plot shows the results assuming the CGGP model and the right plot shows the result assuming the EGP for the MCMC sampling results and the RGGP model for the INLA approximation.

using INLA, were generated in less than 5 seconds on a grid of size 1000.

## 5 DISCUSSION

We show that recent Gaussian process-based Bayesian nonparametric approaches to estimation of effective population size trajectories fall into a larger class of latent Gaussian models, allowing us to perform approximate Bayesian inference using INLA. We show that it is possible to estimate population size trajectories from fixed genealogies in seconds without sacrificing any modeling advantages of recently developed Bayesian nonparametric methods.

We did observe a significant discrepancy between the INLA approximation and MCMC inference for the continuously specified GP model in the case of constant population size. We want to point out that in this case, we are not comparing apples to apples. We should be comparing INLA approximation to the MCMC for the regular grid approximation of the continuously specified GP. However, we did not have access to approximate GP-based MCMC for phylodynamics. In the absence of a better option, we are comparing INLA to the *exact* MCMC for this GP model (Palacios and Minin, 2011). Therefore, we remain uncertain whether the grid approximation or the INLA approximation is to blame for the discrepancy observed in the top plot of Figure 2. The discrepancy between the marginal posterior distributions estimated by INLA and MCMC and the fact that the precision of the RGGP likelihood discretization did not have any effect on our results suggest that INLA approximation indeed fails in this simulation scenario. This assertion is supported by another disagreement of INLA and MCMC for the CGGP model in the influenza A example, where we are comparing apples to apples.

A natural extension of the methods presented here is the incorporation of genealogical uncertainty into the model. This extension can be accomplished by introducing another level of hierarchical modeling and analyzing molecular data directly (Drummond et al., 2005; Minin et al., 2008). Even though the full posterior distribution of population trajectories from molecular sequence data no longer falls into the latent Gaussian model class, we believe that the extension is possible using Metropolis independence sampler (Rue et al., 2004). Nevertheless, the ability to obtain fast estimates of population size trajectories from a fixed genealogy (as with INLA) should be a boon for biological researchers who need to screen multiple populations of interest quickly or to provide an online analysis of epidemic outbreaks with enormous flow of molecular data in real time (Fraser et al., 2009).

There are other approaches to the estimation of effec-tive population sizes under more complicated coalescent models that include recombination (McVean and Cardin, 2005; Li and Durbin, 2011). These methods assume a simple change point model for the effective population size trajectory. In principle, Bayesian nonparametric approaches similar to the approaches discussed here can be applied in this setting. However, presence of recombination makes such extensions potentially challenging.

Other approximate Bayesian methods could be applied to Bayesian nonparametric phylodynamics, such as variational Bayes (VB) (Bishop, 2006) and expectation propagation (EP) (Cseke and Heskes, 2010). For our particular application with a sparse GP prior, such as Brownian motion, Cseke and Heskes (2010) show that INLA should be faster than EP methods.

**ACKNOWLEDGEMENTS**

We acknowledge the `R-INLA` discussion group for helpful comments and the reviewers for their comments and suggestions. This work was supported by the NSF grant No. DMS-0856099. The authors partially completed this research while participating in the Program on Mathematical and Computational Approaches in High-Throughput Genomics at the NSF Institute of Pure and Applied Mathematics, UCLA.